\documentstyle[amsfonts,preprint,aps]{revtex}

\begin{document}
\title{Experimental Purification\ of Single Qubits.}
\author{M. Ricci$^{1}$, F. De Martini$^{1}$, N. J. Cerf$^{2}$, R. Filip$^{3}$, J.
Fiur\'{a}\v{s}ek$^{2,3}$ and C. Macchiavello$^{4}$}
\address{$^{1}$Dipartimento di Fisica and Istituto Nazionale per la Fisica della\\
Materia, Universit\`{a} di Roma ''La Sapienza'', p.le A. Moro 5, Roma\\
I-00185, Italy\\
$^{2}$Ecole Polytechnique, Universit\.{e} Libre de Bruxelles, Bruxelles\\
B-1050, Belgium \\
$^{3}$ Department of Optics, Palack\'{y} University, 17. listopadu 50,\\
Olomouc 77200, Czech Republic \\
$^{4}$ Dipartimento di Fisica ``A. Volta'', Via Bassi 6, Pavia I-27100, Italy}
\maketitle

\begin{abstract}
We report the experimental realization of the purification protocol for
single qubits sent through a depolarization channel. The qubits are
associated with polarization encoded photon particles and the protocol is
achieved by means of passive linear optical elements. The present approach
may represent a convenient alternative to the distillation and error
correction protocols of quantum information.{\bf \ }
\end{abstract}


Modern quantum data processing using realistic (imperfect) quantum gates and
long-distance quantum communication in the presence of a noisy environment
requires a large supply of qubits with a high degree of {\em purity}. Indeed
the fidelity of most quantum information (QI) protocols critically depends
on the preservation of the purity of the QI carriers. It is therefore
crucial to develop techniques that protect quantum states from the
unavoidable losses and decoherence processes accompanying the transmission.
One of these techniques is the quantum error correction \cite
{Shor95,Steane96} which works by encoding the quantum state into a
higher-dimensional Hilbert space. Alternatively, one can distribute several
copies of an entangled state and extract fewer highly-entangled states by
means of entanglement distillation \cite
{Bennett96,Deutsch96,Yamamoto03,Pan03,Zhao03} in order to be able to
subsequently transmit an arbitrary state with high fidelity by quantum
teleportation \cite{Bennett93,Boschi98,Bouwmeester97,Marcikic03}. Yet
another option is to transmit several copies of the state over the noisy
channel and then purify the resulting mixed states at the receiver's station 
\cite{Cirac99,Keyl01,Fischer01}.

The present work realizes the purification procedure that was theoretically
proposed by Cirac {\it et al.} in 1999 \cite{Cirac99}. It addresses the
issue of the purification of $N$ equally prepared qubits in the mixed state $%
\rho =\xi \left| \phi \right\rangle \left\langle \phi \right| +{\frac{1}{2}}%
(1-\xi ){\Bbb I}$, where $0\leq \xi \leq 1$. This procedure allows to
distill from a set of mixed states a subset of states with a higher degree
of purity, i.e. it probabilistically increases the purity by filtering out
some of the noise. The procedure is based on a set of projections onto the
symmetric subspace of the $N$ qubits (i.e. the subspace spanned by all the
states that are invariant under any permutation of the $N$ qubits) and onto
orthogonal subspaces that contain symmetric subspaces for subsets of the
initial $N$ qubits. The procedure is designed to be optimal and universal,
i.e., it acts with the same fidelity for all input states. Since it is
optimal, the purity cannot be further increased by any means. In this paper
we consider the case of two qubits, i.e. $N=2$. The purification procedure
for $N=2$ reduces to a projection of the two-qubit state onto the symmetric
subspace, and it is equivalent to the symmetrization procedure proposed as a
theoretical method to stabilize quantum computation in the presence of noise 
\cite{Bare97}.

For $N=2$ the purification procedure works as follows. Consider two
independent qubits, $a$ and $b$, both originally in the state $\left| \phi
\right\rangle $, that are transmitted over a noisy channel from which they
emerge in a mixed state represented by the density matrix $\rho _{a}=\xi
\left| \phi \right\rangle \left\langle \phi \right| +{\frac{1}{2}}(1-\xi )%
{\Bbb I}=\frac{1+\xi }{2}\left| \phi \right\rangle \left\langle \phi \right|
+\frac{1-\xi }{2}\left| \phi ^{\bot }\right\rangle \left\langle \phi ^{\bot
}\right| $ , where $|\phi ^{\bot }\rangle $ is a state orthogonal to $|\phi
\rangle $. Our goal is then to purify the transmitted qubits in order to
obtain two qubits that are as close as possible to the original state $%
\left| \phi \right\rangle $. The overall 2-qubit input state $\rho
_{ab}^{in}=\rho _{a}^{in}\otimes \rho _{b}^{in}$ is expressed in the basis $%
\{\left| \phi \right\rangle _{a}\left| \phi \right\rangle _{b},\left| \phi
\right\rangle _{a}\left| \phi ^{\perp }\right\rangle _{b},\left| \phi
^{\perp }\right\rangle _{a}\left| \phi \right\rangle _{b},\left| \phi
^{\perp }\right\rangle _{a}\left| \phi ^{\perp }\right\rangle _{b}\}$ by the
matrix 
\begin{equation}
\rho _{ab}^{in}=\frac{1}{4}\left( 
\begin{array}{cccc}
\left( 1+\xi \right) ^{2} & 0 & 0 & 0 \\ 
0 & 1-\xi ^{2} & 0 & 0 \\ 
0 & 0 & 1-\xi ^{2} & 0 \\ 
0 & 0 & 0 & \left( 1-\xi \right) ^{2}
\end{array}
\right)
\end{equation}
As mentioned above, the purification protocol consists of the projection of
the 2-qubit state onto the symmetric subspace: if the projection is
successful we obtain two equal output qubits that are the optimal
''purified'' ones, otherwise we discard the output states. We note that this
protocol can be implemented, for every qubit encoding, by a quantum circuit
requiring an ancilla qubit and a Toffoli gate \cite{Scia04}. After a
successful projection the output qubits are in the state 
\begin{equation}
\rho _{ab}^{out}=\frac{\Pi \rho _{ab}^{in}\Pi ^{\dagger }}{Tr[\Pi \rho
_{ab}^{in}\Pi ^{\dagger }]}=\frac{1}{3+\xi ^{2}}\left( 
\begin{array}{cccc}
\left( 1+\xi \right) ^{2} & 0 & 0 & 0 \\ 
0 & \frac{1-\xi ^{2}}{2} & \frac{1-\xi ^{2}}{2} & 0 \\ 
0 & \frac{1-\xi ^{2}}{2} & \frac{1-\xi ^{2}}{2} & 0 \\ 
0 & 0 & 0 & \left( 1-\xi \right) ^{2}
\end{array}
\right)
\end{equation}
where $\Pi ={\Bbb I}_{ab}-|\Psi _{ab}^{-}\rangle \langle \Psi _{ab}^{-}|$ is
the projector onto the symmetric subspace and $|\Psi _{ab}^{-}\rangle
=(|01\rangle -|10\rangle )/\sqrt{2}$ is the singlet state of two qubits. The
success probability of the procedure is $p=Tr[\Pi \rho _{ab}^{in}\Pi
^{\dagger }]=\frac{3+\xi ^{2}}{4}$. Since $\rho _{ab}^{out}$ belongs to the
symmetric subspace, the reduced density matrices of the resulting single
qubits, expressed in the basis $\{\left| \phi \right\rangle ,\left| \phi
^{\perp }\right\rangle \}$, are found to be identical, 
\begin{equation}
\rho _{a}^{out}=\rho _{b}^{out}=\frac{1}{2}\left( 
\begin{array}{cc}
1+\xi _{P} & 0 \\ 
0 & 1-\xi _{P}
\end{array}
\right) \ ,
\end{equation}
where $\xi _{P}=\frac{4}{3+\xi ^{2}}\xi $ $\geq \xi $ and the purification
gain factor is $\eta =\frac{4}{3+\xi ^{2}}.$ Note that $p$ and $\eta $ are
related by the equation $\eta p=1$ so a higher purification gain factor is
necessarily accompanied by a lower probability of success.

We report the implementation of the above protocol for qubits encoded in the
polarization of single photons (see Figure 1). The qubit to be purified is $%
\frac{1+\xi }{2}\left| \phi \right\rangle \left\langle \phi \right| +\frac{%
1-\xi }{2}\left| \phi ^{\perp }\right\rangle \left\langle \phi ^{\perp
}\right| $ where $\left| \phi \right\rangle =\alpha \left| H\right\rangle $ $%
+\beta \left| V\right\rangle $ and $\left| H\right\rangle $, $\left|
V\right\rangle $ respectively correspond to the horizontal and vertical
linear polarizations. In the present experiment, pairs of photons with
wavelength $\lambda =532$~nm and coherence time $\tau _{coh}=80$~fs, were
generated in a Type I, BBO\ crystal slab in the {\it product state} $\left|
H\right\rangle _{a}\left| H\right\rangle _{b}$ by spontaneous parametric
down conversion (SPDC) process excited by a CW fourth-harmonic-generation
laser (Coherent \ Verdi +MBD-266).\ \ The output state was first encoded in
the state $\left| \phi \right\rangle _{a}\left| \phi \right\rangle _{b}$ by
means of two {\it equal }waveplates (wp) $WP(\left| \phi \right\rangle )$
and then each photon, injected into a noisy channel $P$, emerged in the
mixed state: $\rho _{a}=\rho _{b}$ $=\xi \left| \phi \right\rangle
\left\langle \phi \right| +(1-\xi )\frac{{\Bbb I}}{2}$. The two mixed
qubits, associated with the two modes $k_{a}$ and $k_{b}$, were linearly
superimposed at beam-splitter $BS$ with a mutual time delay $\Delta t$
micrometrically adjustable by a translation stage with position settings $%
Z=2\Delta tc$, with $c$ denoting the velocity of light. The value $Z=0$ was
assumed to correspond to the full overlapping of the photon pulses injected
into $BS$, i.e. to the maximum photon interference leading to the
simultaneous detection of two photons on either output modes $k_{1}$ or $%
k_{2}$ of $BS$ \cite{Hend03}\cite{Ricc04}. Recently it has been shown that
the projection of the overall state in the symmetric subspace, precisely the
one implying the present purification procedure, is unambiguously identified
by the maximum interference condition: $Z=0$ \cite{Ricc04}\cite{Irvi04}.

Let us give more details about the realization of the two equal depolarizing
channels $P$ and $P^{\prime }$ operating on the $BS$ input modes $k_{a}$, $%
k_{b}$, respectively. Each channel implemented the quantum map ${\cal E}%
\left( \rho \right) =\xi \rho +(1-\xi ){\cal E}_{DEP}\left( \rho \right) $
where ${\cal E}_{DEP}\left( \rho \right) $ maps any unknown input state $%
\rho $ into a {\it fully mixed} one. This transformation can be achieved by
stochastically applying the full set of Pauli operators \{${\Bbb I},$ $%
\sigma _{x},\sigma _{y},\sigma _{z}$\} with the same statistical weight,
that is, ${\cal E}_{DEP}\left( \rho \right) ={\frac{1}{4}}({\Bbb I}\rho 
{\Bbb I}+\sigma _{x}\rho \sigma _{x}+\sigma _{y}\rho \sigma _{y}+\sigma
_{z}\rho \sigma _{z})$. Let us consider here only one of the noisy channels,
say $P$, the one operating on the mode $k_{a}$. The ${\cal E}\left( \rho
\right) $ map was realized by means of a pair of equal electro-optic $(EO)$
LiNbO$_{3}$ Pockels cells (P-cell), $P_{X}$ and $P_{Z}$, carefully aligned
with a $45^{\circ }$ mutual spatial orientation of the optical axes of the
EO crystals (see Fig. 1). All P-cells were Shanghai Institute of Ceramics
devices with a $\frac{\lambda }{2}$-voltage: $V_{\frac{\lambda }{2}}=390$~V.
Each P-cell of the pair was driven by a CW periodic square-wave electric
field with maximum $V=V_{\frac{\lambda }{2}}$, {\it fixed} frequency $f=T$ $%
^{-1}$ and {\it variable} pulse duration $\tau $ corresponding to a {\it %
duty-cycle }$\nu =\tau /T$ adjustable in the range: $0<v<1/2$. The
excitation pulses feeding the two P-cells were mutually delayed by a time
equal to $\tau /2$ (see inset of Fig. 1). Consider a single excitation
cycle. In the time intervals $\Delta \tau =\tau /2$ in which only one P-cell
was active, either the $\sigma _{x}$ or the $\sigma _{z}$ transformation was
implemented\ depending on the corresponding crystal orientation. In the
interval $\Delta \tau =\tau /2$ in which both P-cells were simultaneously
active, the $\sigma _{y}$ transformation was realized. In summary, each
operators ${\Bbb I},\sigma _{x},\sigma _{y},\sigma _{z}$ was applied to the
input state over a time $\Delta \tau =\tau /2$ and the total depolarizing
process lasted a time $2\tau $ over each period $T$, thus achieving an
average depolarizing fraction $(1-\xi )=2\nu $. In order to avoid any
correlation between the two qubits to be purified, $\rho _{a}$ and $\rho
_{b} $, the two channels $P$ and$\ P^{\prime }$ were driven by different
frequencies: $f$ =6 KHz and $\ f^{\prime }=1.7\times f$. Correspondingly,
different values of $\tau $ were adopted for the two channels in order to
realize, within each experimental run, equal values of $\xi $ for the two
input qubits.

In the analysis we have assumed an identical preparation of the two input
qubits, while the output ones are described by the same density matrix $\rho
=\rho _{a}=\rho _{b}$. With this assumption, carefully checked over each
channel, the verification of the purification procedure lies on the
tomography of the density matrix of one of the input and one of the output
qubits. For the sake of simplicity, we only analyzed the measurements
performed on the $BS$ output mode $k_{1}$ (see Fig. 1), selecting counts in
coincidence between the detectors $[D_{1},D_{2}]$ to trigger the realization
of the projection of $\rho $ onto the symmetric sub-space. The detectors $%
D_{1},D_{2}$ were coupled to mode $k_{1}$ by a 50:50 beam-splitter $BS_{1}$. 
$D_{1}$ provided the measured outcomes of a simple tomographic setup
consisting of a $\lambda /2$-wp, a $\lambda /4$-wp and a polarizing beam
splitter ($PBS$).

Consider first the projector switched off, by setting $Z\gg c\tau _{coh}$,
i.e., by spoiling any interference on the photons impinging on $BS$. A
tomographic reconstruction of the qubit in the mode $k_{1}^{^{\prime }}$
based on the measurement of the corresponding Stokes parameters by 4
different settings of the wp's $\frac{\lambda }{2},\frac{\lambda }{4}$ was
undertaken. It is easy to see that this qubit, corresponding to the qubit to
be purified, is expressed by the density matrix $\rho _{1}^{^{\prime }}=%
\frac{\rho _{a}^{in}+\rho _{b}^{in}}{2}$ \cite{Note}. By turning on the
projector, i.e., by restoring the $BS$ interference setting $Z=0$, the mode $%
k_{1}$ contains the two photons described by the density matrix $\rho
_{ab}^{out}$. In this case, we measured on the mode $k_{1}^{^{\prime }}$ the
purified qubit $\rho _{1}^{^{\prime }}=$ $\rho _{a}^{out}.$ From the density
matrices reconstructed in absence and in presence of interference, we obtain 
$\xi $, $\xi _{p}$ and thus the purification factor $\eta =\xi _{p}/\xi $.
In addition, from the coincidence rates determined for $Z=0$ and for $Z\gg
c\tau _{coh}$ we inferred the success probability $p$ of the purification
protocol. We may check that an increase of the purification gain factor for
any qubit pair, i.e., a larger $\eta $, corresponds to a lower success
probability of the overall protocol, as expected. In Fig. 2, we plotted the
experimental values of $\eta $ and $p$ obtained for different $\xi $'s,
i.e., for different experimental values of $\nu =(1-\xi )/2$, for three
input states: $\left| H\right\rangle $, $\left| L\right\rangle =2^{-{\frac{1%
}{2}}}(\left| H\right\rangle +\left| V\right\rangle )$, and $\left|
E\right\rangle =[\cos (3\pi /16)\left| H\right\rangle +i\sin (3\pi
/16)\left| V\right\rangle ]$ corresponding, respectively, to horizontal, $%
45^{\circ }$-diagonal, and a very general elliptical polarizations of the
input qubits. The mutual agreement of the data for different input states
demonstrates the universality of the purification procedure. The deviations
of the experimental data from the theoretical values were mainly due to the
imperfections of the optical components. In particular, the non-ideal
properties of the main $BS$ were found to be highly critical. In order to
achieve the projection onto the symmetric subspace, the $BS$ transmittances $%
T_{H}$ and $T_{V}$ for the $H$ and $V$ polarization modes should be equal,
with a high level of precision, and any difference between $T_{H}$ and $%
T_{V} $ partially spoils the purification. Notice, however, that deviations
of $T_{H}=T_{V}$ from 50\% only decrease the success probability $p$ but do
not alter the purification gain factor $\eta $.

We may generalize the above method by accounting for any possible asymmetry
of the preparation of the input qubits. Allowing the input qubits to have
different degree of mixedness, i.e. $\rho _{a}^{in}=\zeta \left| \phi
\right\rangle \left\langle \phi \right| +(1-\zeta )\frac{I}{2},$\ $\rho
_{b}^{in}=\kappa \left| \phi \right\rangle \left\langle \phi \right|
+(1-\kappa )\frac{I}{2}$, the output qubits are still in the state given by
Eq.3 with $\xi _{P}$=$2(\zeta +\kappa )/(3+\kappa \zeta )$. The purification
factor is $\eta =\xi _{P}/\xi =\frac{4}{3+\kappa \zeta }=1/p$ where $\xi
\equiv 
{\frac12}%
(\zeta +\kappa )$ is the average input{\em \ }mixedness factor. This process
may be investigated by recourse to the quantum ''relative entropy'' that
measures the closeness of any output state $\sigma $ with respect to a
corresponding input pure state $\rho $, e.g. after propagation through a
noisy communication channel: $S(\rho \parallel \sigma )\equiv Tr(\rho
Log\rho )-$ $Tr(\rho Log\sigma )\;$\cite{Niel00}. Suppose that two qubits
are equally prepared in the {\it pure} state $\rho =$ $\left| \phi
\right\rangle _{a}\left\langle \phi \right| _{a}\otimes \left| \phi
\right\rangle _{b}\left\langle \phi \right| _{b}$, $S(\rho )=0$ . After
corruption by noise the entropy is: $S(\rho \parallel \rho ^{in})$=$Log\left[
{\frac12}%
(1+\zeta )\right] +Log\left[ 
{\frac12}%
(1+\kappa )\right] $. If the qubits are further purified by symmetrization
the following result is obtained: $S(\rho \parallel \rho ^{out})$=$Log\left[ 
{\frac12}%
(1+\zeta )\right] +Log\left[ 
{\frac12}%
(1+\kappa )\right] -Log\eta $. Then the symmetrization leads to the positive 
{\it information gain} $\Delta S$=$S(\rho \parallel \rho ^{in})-S(\rho
\parallel \rho ^{out})$ = $Log\eta $ \ at the expense of a reduced rate $p$
of success: $\Delta S=-Logp$, $%
{\frac34}%
\leq p\leq 1$.

An interesting case is represented by the purification of a\ {\it fully} 
{\it mixed state} by a {\it pure} state, e.g. by the initial conditions $%
\zeta $=$1$ and $\kappa $=$0$. This precisely corresponds to the
probabilistic {\it quantum cloning} process recently realized by our
Laboratory in Rome by a symmetrization procedure \cite{Scia04},\cite{Ricc04}%
. There $\xi =%
{\frac12}%
$, and a purification gain factor $\eta =4/3$ was attained with a success
probability: $p=3/4$. \newline

In conclusion, we have experimentally demonstrated the optimal purification
of two depolarized qubits using the interference of two photons at a beam
splitter, conditionally effecting symmetrization. The experimentally
observed purification gain factors are in very good agreement with the
theoretical estimates. We therefore envision that single qubit purification
may become a viable procedure for protecting quantum states against noise.
Note also that the projection on the symmetric subspace of more than two
qubits can be carried out with a sequence of beam splitters, which is in the
reach of present optical technology.

MR and FDM acknowledge financial support from the FET European Network
IST-2000-29681 (ATESIT), the Istituto Nazionale per la Fisica della Materia
(PRA-CLON) and the Ministero della Istruzione, dell'Universita' e della
Ricerca (COFIN 2002). NC and JF acknowledge financial support from the
Communaut\'{e} Francaise de Belgique, from the IUAP programme of the Belgian
govrnment and from the EU under project CHIC. RF and JF also acknowledge
support from the Czech Ministry of Education under the project Research
Center for Optics. CM acknowledge financial support from the FET European
Network IST-ATESIT and IST-QUPRODIS

\centerline{\bf Figure Captions}

\vskip 8mm

\parindent=0pt

\parskip=3mm

Figure 1. Experimental setup for the optical implementation of the
purification procedure. Inset shows the realization of the depolarizing
channel employing two Pockels cells.

Figure 2. Experimental results of the purification procedure for different
input qubits corresponding to the encoded polarizations: $\left|
H\right\rangle $, $\left| L\right\rangle =2^{-1/2}\left( \left|
H\right\rangle +\left| V\right\rangle \right) $, and $\left| E\right\rangle
=\left( \cos \left( \frac{\theta }{2}\right) \left| H\right\rangle +i\sin
\left( \frac{\theta }{2}\right) \left| V\right\rangle \right) $ with $\theta
=\frac{3}{8}\pi $. Filled markers denote the experimental purification
factor $\eta $ data while open markers denote the experimental data of the
procedure probability $p$. For simplicity we report only one error-bar for
the probability values obtained. The statical error is the same for all data
reported.

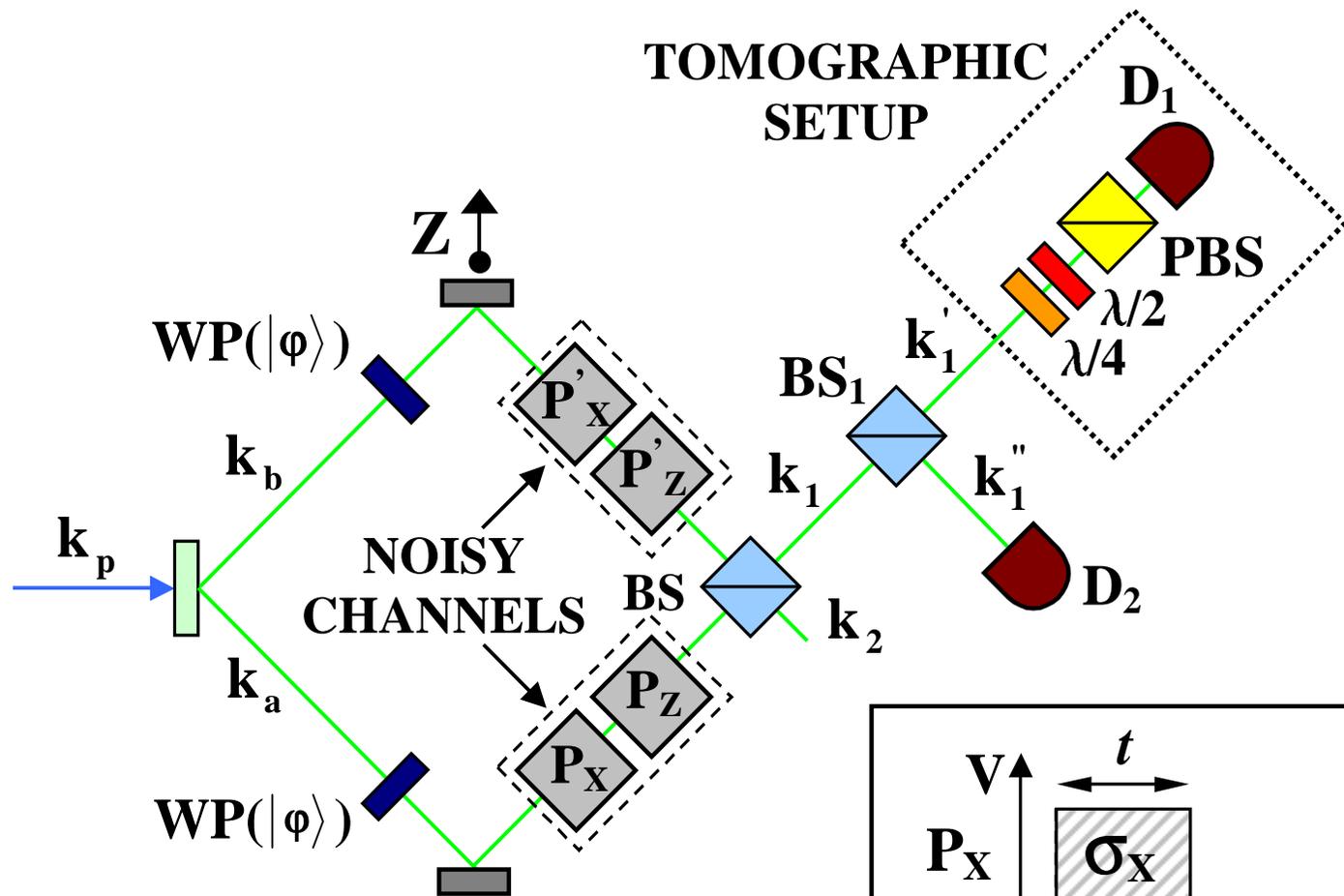
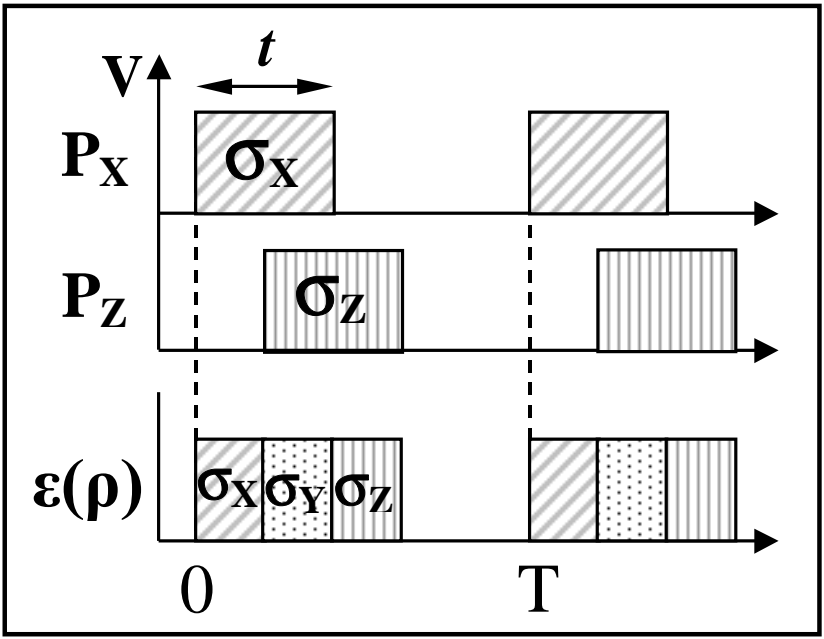
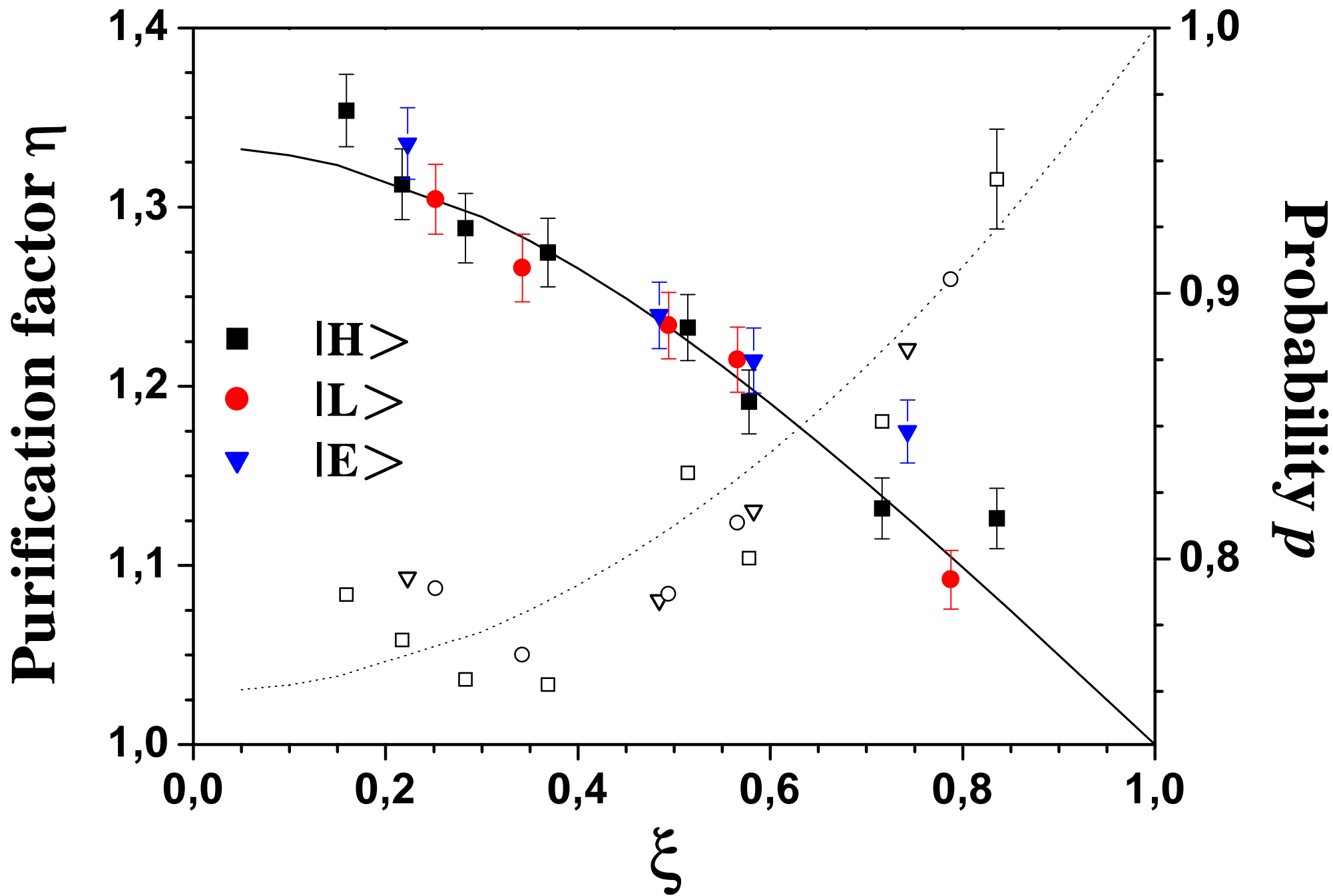
\end{document}